\documentclass[iop]{emulateapj}
\usepackage{float}
\usepackage{rotating} 
\usepackage{times}
\usepackage{subfigure}
\usepackage{amsmath}
\usepackage{amsfonts}
\usepackage{amssymb}
\usepackage{epsfig}
\usepackage{hyperref}
\usepackage{graphicx}
\usepackage{natbib}

\usepackage{bm}
\usepackage{color}
\usepackage{hyperref}

\usepackage{amsmath}
\usepackage{amsfonts}
\usepackage{amssymb}
\usepackage{epsfig}
\usepackage{hyperref}
\usepackage{graphicx}

\begin{document}

\title{HerMES: The Rest-Frame UV Emission and A Lensing Model for the $\lowercase{z}=6.34$ Luminous Dusty Starburst Galaxy HFLS3}

\author{ 
Asantha~Cooray\altaffilmark{1}, 
Jae~Calanog\altaffilmark{1}, 
Julie~L.~Wardlow\altaffilmark{2},
J.~Bock\altaffilmark{3,4},
C.~Bridge\altaffilmark{3},
D.~Burgarella\altaffilmark{5},
R.~S.~Bussmann\altaffilmark{6},
C.~M.~Casey\altaffilmark{1}, 
D.~Clements\altaffilmark{7}, 
A.~Conley\altaffilmark{8}, 
D.~Farrah\altaffilmark{9}, 
H.~Fu\altaffilmark{10},
R.~Gavazzi\altaffilmark{11},
R.~J.~Ivison\altaffilmark{12,13},
N.~La Porte\altaffilmark{14},
B.~Lo~Faro\altaffilmark{15},
Brian~Ma\altaffilmark{1},
G.~Magdis\altaffilmark{16},
S.~J.~Oliver\altaffilmark{17},
W.~A.~Osage\altaffilmark{1},
I.~P{\'e}rez-Fournon\altaffilmark{18},
D.~Riechers\altaffilmark{6},
D.~Rigopoulou\altaffilmark{16},
Douglas~Scott\altaffilmark{19}
M.~Viero\altaffilmark{1},
D.~Watson\altaffilmark{2}
}

\altaffiltext{1}{Center for Cosmology, Department of Physics  and  Astronomy, University  of California, Irvine, CA 92697, USA}
\altaffiltext{2}{Dark Cosmology Centre, Niels Bohr Institute, University of Copenhagen, Juliane Maries Vej 30, 2100 Copenhagen, Denmark}
\altaffiltext{3}{California Institute of Technology, 1200 E. California Blvd., Pasadena, CA 91125, USA}
\altaffiltext{4}{Jet Propulsion Laboratory, 4800 Oak Grove Drive, Pasadena, CA 91109, USA}
\altaffiltext{5}{Laboratoire d'Astrophysique de Marseille,  Aix-Marseille University,  CNRS, 13013 Marseille, France}
\altaffiltext{6}{Department of Astronomy, Space Science Building, Cornell University, Ithaca, NY, 14853-6801, USA}
\altaffiltext{7}{Astrophysics Group, Imperial College London, Blackett Laboratory, Prince Consort Road, London SW7 2AZ, UK}
\altaffiltext{8}{Dept. of Astrophysical and Planetary Sciences, CASA 389-UCB, University of Colorado, Boulder, CO 80309, USA}
\altaffiltext{9}{Department of Physics, Virginia Tech, Blacksburg, VA 24061, USA}
\altaffiltext{10}{Department of Physics \& Astronomy, University of Iowa, Iowa City, IA 52242, USA}
\altaffiltext{11}{Institut d'Astrophysique de Paris, UMR 7095, CNRS, UPMC Univ. Paris 06, 98bis boulevard Arago, F-75014 Paris, France}
\altaffiltext{12}{European Southern Observatory, Karl Schwarzschild Strasse 2, D-85748 Garching, Germany}
\altaffiltext{13}{Institute for Astronomy, University of Edinburgh, Royal Observatory, Blackford Hill, Edinburgh EH9 3HJ, UK}
\altaffiltext{14}{Pontificia Universidad Ca\'olica de Chile, Departamento de Astronom{\'\i}a y Astrof{\'\i}sica, Casilla 306, Santiago 22, Chile}
\altaffiltext{15}{Aix-Marseille Universitè, CNRS, LAM (Laboratoire d’Astrophysique de Marseille) UMR7326, 13388, France}
\altaffiltext{16}{Department of Astrophysics, Denys Wilkinson Building, University of Oxford, Keble Road, Oxford OX1 3RH, UK}
\altaffiltext{17}{Astronomy Centre, Dept. of Physics \& Astronomy, University of Sussex, Brighton BN1 9QH, U}
\altaffiltext{18}{Instituto de Astrof{\'\i}sica de Canarias (IAC), E-38200 La Laguna, Tenerife, Spain}
\altaffiltext{19}{Department of Physics \& Astronomy, University of British Columbia, 6224 Agricultural Road, Vancouver, BC V6T~1Z1, Canada}


\begin{abstract} 
We discuss the rest-frame ultraviolet emission from the starbursting galaxy HFLS3 at a redshift of 6.34. 
The galaxy was discovered in {\it Herschel}/SPIRE data due to its {\it red} color in the sub-mm wavelengths from 250 to 500 $\mu$m.
The apparent instantaneous star-formation rate  of HFLS3 inferred from the total far-IR luminosity measured with
over 15 photometric data points between 100 and 1000 $\mu$m is 2900 M$_{\odot}$ yr$^{-1}$.
Keck/NIRC2 K$_s$-band adaptive optics imaging data showed two 
potential near-IR counterparts near HFLS3. Previously, the northern galaxy was taken to be in the foreground at $z=2.1$ while
the southern galaxy was assumed to HFLS3's  near-IR counterpart.
The recently acquired {\it Hubble}/WFC3 and ACS imaging data show conclusively that both optically-bright galaxies are 
in the foreground at $z < 6$. A new lensing model based on the {\it Hubble} imaging data 
and the mm-wave continuum emission yields a magnification factor of $2.2 \pm 0.3$.  The lack of multiple imaging constrains the lensing magnification
to be lower than either 2.7 or 3.5 at the 95\% confidence level for the two scenarios, which attribute
one or two components to HFLS3 in the source plane.
Once accounting for the possibility of gravitational lensing, the instantaneous star-formation rate is 1320 M$_{\sun}$ yr$^{-1}$ with the 95\% confidence
lower limit around 830 M$_{\sun}$ yr$^{-1}$. Using models for the rest-frame UV to far-IR spectral energy distribution (SED)
we determine the average star-formation rate over the last 100 Myr  to be around 660 M$_{\odot}$ yr$^{-1}$.
The dust and stellar masses of HFLS3 from the same SED models are at the level of $3 \times 10^8$ M$_{\odot}$ and $\sim 5 \times 10^{10}$ M$_{\odot}$, respectively, with
large systematic uncertainties on assumptions related to the SED model. With {\it Hubble}/WFC3 images we also find diffuse near-IR emission about 0.5 arcseconds ($\sim$ 3 kpc)
 to the South-West of HFLS3 that remains undetected in the ACS imaging data. The emission has a photometric redshift consistent with either $ z\sim 6$ or a dusty galaxy
template at $ z\sim 2$. If at the same redshift as HFLS3 the detected diffuse emission could be
 part of the complex merger system that could be triggering the starburst. Alternatively, it could be part of the foreground structure at $z \sim 2.1$ that is
responsible for lensing of HFLS3.
\end{abstract}


\section{Introduction}
The unexpected discovery of HFLS3 (HerMES J170647.8+584623) at a redshift of $6.3369 \pm 0.0009$ 
in {\it Herschel} Space Observatory's \citep{Pilbratt10} has
led to the possibility that massive starbursting galaxies could 
be an appreciable contributor to the star-formation rate density of the Universe during the epoch of reionization \citep{Riechers13}.  
The galaxy was first identified in  {\it Herschel} Multi-Tiered Extragalactic Survey (HerMES\footnote{http://hermes.sussex.ac.uk}, \citealt{Oliver12})
as a high-redshift candidate due to its ``red'' color in the
SPIRE  \citep{Griffin10} data, with $S_{500}/S_{350} \sim 1.45$ and $S_{500} \sim 47 \pm 3$ mJy.
The redshift of HFLS3 was secured through the
detection of more than 20 individual molecular and atomic lines at far-IR/sub-mm wavelengths with ground-based interferometers.
HFLS3 was found to be luminous ($L_{\rm IR}=(3.4 \pm 0.3)\times10^{13}$ L$_{\odot}$),
gas-rich ($M_{\rm gas} \sim 10^{11}$ M$_{\odot}$) and dusty ($T_{\rm d}=49\pm2$ K). 
The instantaneous star-formation rate (SFR) implied by the above total IR luminosity
\citep{Kennicutt98}  is around 2900 M$_{\odot}$ yr$^{-1}$ for a \citet{Chabrier03} initial mass function.
It is also the highest redshift sub-mm galaxy (SMG) known to date, potentially probing the earliest formation epoch of dust in the Universe
\citep*[for a recent review of SMGs and dusty star-forming galaxies in general see][]{Casey14}.

One complication in interpreting the properties of HFLS3 is that it was found to be $\sim 0.5\arcsec$ to the South of
a $z= 2.09$ galaxy (Figure~1), identified by Keck/NIRC2 K-band AO imaging and Keck/LRIS spectroscopy.
This suggests some possibility that the flux density of HFLS3 is enhanced by gravitational lensing with a magnification factor, $\mu_{\rm lens}$.  
Due to the steepness of the SMG  number counts and their high redshifts,  and the corresponding high magnification bias,
sub-mm surveys are known to be highly sensitive to gravitational lensing modifications \citep{Blain96,Perrotta02,Negrello07,Paciga09}. At the bright-end of the number counts at wavelengths
longer than 350 $\mu$m, lensed SMGs appear as a power-law  distinct from the intrinsic counts (e.g., \citealt{Negrello10,Wardlow13,Vieira13}).
At $z > 4$, we expect the lensing fraction to be substantial for current generation surveys, where the flux density limit for the source detection is relatively high.
An example of a high efficiency lensing selection at $z > 3$ is the  bright SMG sample from the South Pole Telescope at 1.4 mm \citep{Vieira13,Weiss13}. 
If lensing is a statistically important correction to the
flux densities of high-redshift SMGs we expect them to be discovered near foreground galaxies and groups. Such a close association with a foreground galaxy 
is consistent with the existing indications that a reasonable fraction of the $z > 7$ Lyman-break drop-outs are
also magnified by  $\mu_{\rm lens}\sim$ few due to their closeness to foreground bright galaxies \citep{Wyithe11}.

In the case of HFLS3, a possibility for lensing was expected since the Keck/LRIS spectroscopy showed emission lines corresponding to a foreground galaxy at
a $z=2.1$ within one arcsecond of the  peak 1.1 mm continuum emission. 
The high resolution Keck/NIRC2 LGS-AO imaging data in the K$_s$-band showed two galaxies within $1\farcs5$ of HFLS3. 
In \citet{Riechers13} the northern component was taken 
to be the $z=2.1$ foreground galaxy, while the southern component, close to the peak 1.1 mm emission, was taken to be the rest-frame optical counterpart, or the
least obscured part, of HFLS3. 
Under such an assumption, deblended NIRC2 and {\it Spitzer}/IRAC photometry suggested a stellar mass of $\sim 3.7 \times 10^{10}$ M$_{\sun}$. Thus, HFLS3 is
already a stellar mass-rich galaxy at $z=6.34$, while continuing to form stars at a very high rate of $>2000$ M$_{\odot}$ yr$^{-1}$.

The lack of multiple images of HFLS3 in mm-wave interferometric imaging data was inferred to imply that the lensing magnification factor is negligible, with
 $\mu_{\rm lens}=1.5 \pm 0.7$ associated with lensing by the foreground galaxy to the north of the assumed rest-frame optical counterpart.  Due to such a small magnification
 a lensing correction to the properties of HFLS3 was not included in \citet{Riechers13}. However,
the lensing magnification determination is subject to assumptions related to the counterpart identification and the location of
foreground galaxies relative to the mm-wave emission.
Since the true mass and star-formation rate of HFLS3 are directly related to its cosmic rarity,  a potential lensing correction is even more important
when addressing whether HFLS3 is a rare source among the SMG sample or if it is a source typical of $z > 4$ SMGs \citep{Daddi09,Coppin10,Capak11,Walter12,Combes12}.

Here we report {\it Hubble}/WFC3 and ACS imaging observations of HFLS3 in five filters from optical to near-IR wavelengths.
 We use these data to study the physical properties of HFLS3 by improving the lensing model and by identifying rest-frame optical/UV emission
for a new estimate of the stellar mass of HFLS3. This paper is organized as follows. In the next Section we summarize the observations and the analysis. We 
discuss the counterpart identification and {\sc Galfit} (Peng et al. 2002) models in
Section~3. Our lens models and the magnification factor of HFLS3 are presented in Section~4. In Section~5 we present the modeling of optical to IR SED  of foreground
galaxies  and the UV to far-IR SED of HFLS3. We present  a discussion of 
our key results and the implications for the presence of massive, dusty starbursts
galaxies at high redshifts in Section~6 and conclude with a summary in Section~7.
For lensing and SED models we assume the best-fit concordance cosmology consistent with 
WMAP-9 year and Planck data \citep{Hinshaw13,Planck13}. 

\begin{figure*}[htb]
\centerline{
\includegraphics[scale = 0.4]{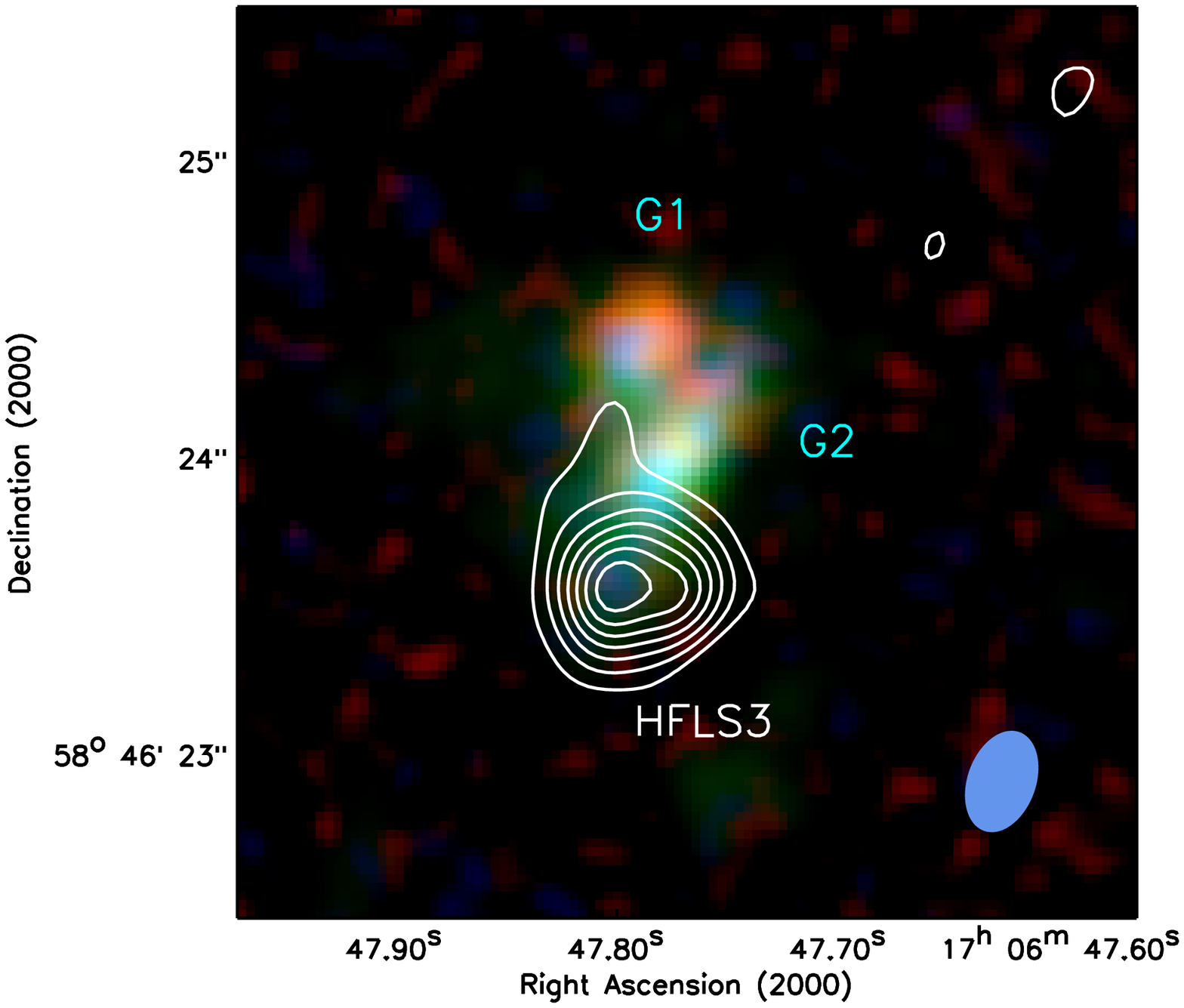}
\includegraphics[scale = 0.4]{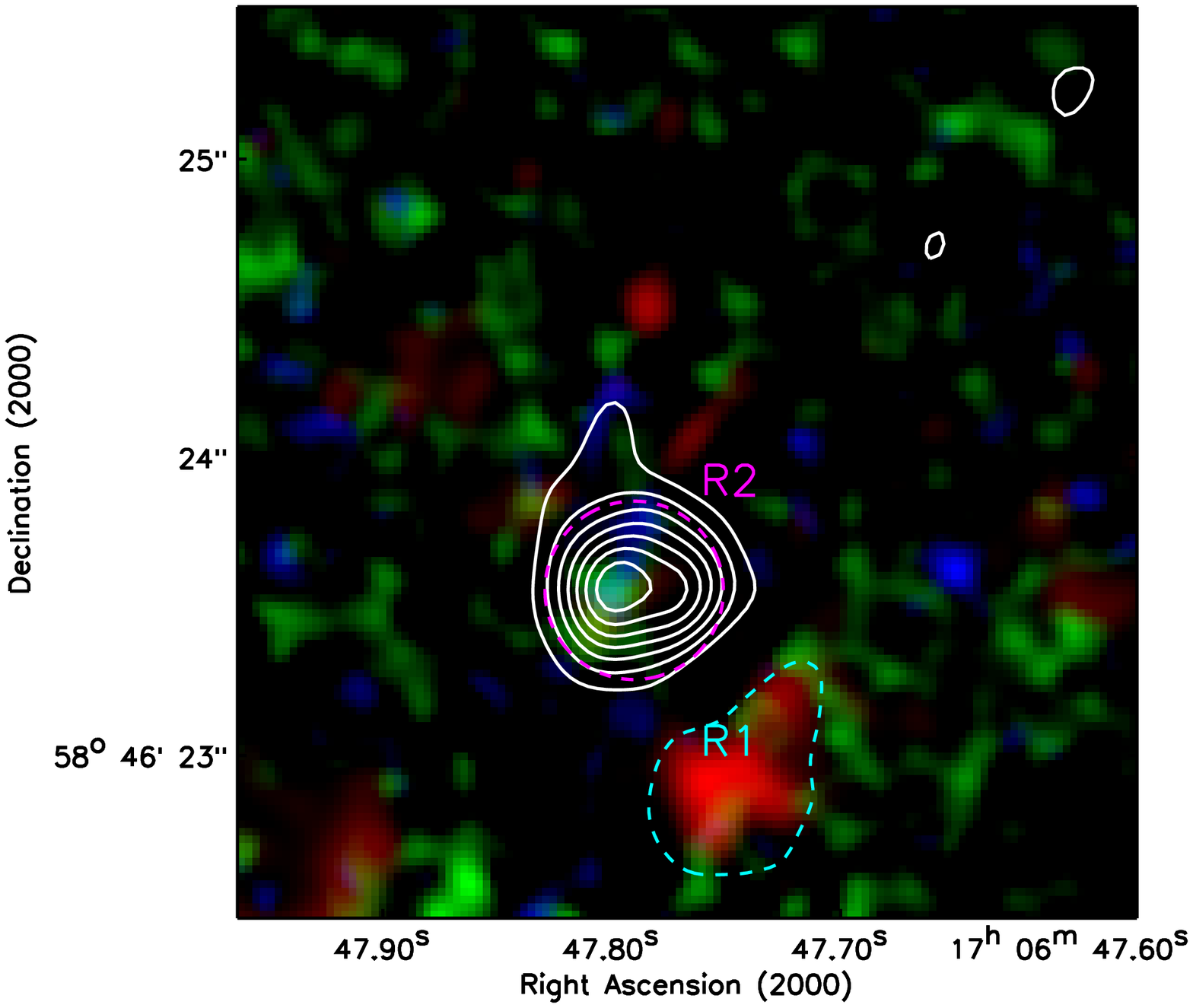}
}
\caption{\label{fig:galfit} 
{\it Left:} The three-color image using HST/ACS combined F625W and F814W (blue), HST/WFC3-IR combined F160W, F125W and F105W, and Keck/NIRC-2 $K_s$-band LGS-AO (red) images.
Note the clear detection of two galaxies close to HFLS3 shown here in terms of the IRAM/PdBI 1.1 mm (rest-frame 158 $\mu$m) emission. The r.m.s. uncertainty 
in the PdBI A-array configuration data is 180 $\mu$Jy beam$^{-1}$ and the contours are shown in steps of 3$\sigma$ starting at 5$\sigma$. The instrumental beam is
shown to the bottom right with FWHM of 0.35$''$ $\times$ 0.23$''$.
{\it Right:} The three-color {\sc GALFIT} residual map where we remove models for the HST/ACS-detected galaxies in HST/WFC3. Here we show the combination of
ACS/F625W$+$F814W (blue), WFC3/F105W (green) and WFC3/F160W (red).
Both G1 and G2 are detected in the combined ACS/F625W and F814W stack, consistent with the scenario that both G1 and G2 are at $z < 6$ and G2 is
not the least obscured region, or the rest-frame optical counterpart, of HFLS3, 
as was previously assumed. We find a marginal detection of rest-UV emission at the location of HFLS3 (labeled R2)
and a higher significance diffuse emission $0\farcs5$ to the South-West of HFLS3 (labeled R1). We use WFC3 fluxes and ACS upper limits of R2 for
combined SED modeling of HFLS3 with far-IR/sub-mm flux densities. We detemine a photometric redshift for 
R1 and find it to be consistent with emission from either a galaxy at $z \sim 6$ or a dusty galaxy at $z\sim 2$.
}
\end{figure*}

\section{Hubble Space Telescope Observations}
\label{sec:data}

HFLS3 was observed with {\it Hubble}/ACS and WFC3 in Cycle 21 (GO 13045; PI Cooray)  in order to understand the nature and environment
of currently the highest redshift dusty starburst known from sub-mm survey data.
The observations were carried out in F160W/F125W/F105W filters with WFC3 and in F814W and F625W with ACS  over a total of six orbits.
The imaging data reach 5$\sigma$ point source depths of $m_{\rm AB} = 26.0$, 26.3, 25.9, 27.0, and 26.1
in F160W, F125W, F105W, F814W, and F625W, respectively.
While the WFC3 imaging was aimed at detecting the rest-frame UV emission from HFLS3,  the ACS imaging was aimed at 
establishing the exact location, size, and morphology of the nearby  $z \sim 2.1$ galaxy for an improved lens model. 
The five band photometry was aimed at completing the rest-UV SED of HFLS3 to improve the stellar mass estimate once
combined with Keck/NIRC2 K$_s$ and {\it Spitzer}/IRAC photometry. Here we focus on properties of HFLS3, but another study will
discuss the environment of HFLS3 (La Porte et al. in preparation). The HST data are also useful for a near-IR counterpart search of
SCUBA-2 sources detected in the HFLS3 field \citep{Robson14}.

The {\it Hubble} data were analyzed with the standard tools. For WFC3 imaging data, we make use of {\sc calwfc3} in the {\sc IRAF.STSDAS} pipeline for
flat-fielding and cosmic-ray rejection. Individual exposures in each of the filters were combined with {\sc Astrodrizzle} \citep{Fruchter10}
and we produced images at a pixel scale of $0\farcs06$ from the native scale of $0\farcs13$ per pixel. For flux calibration we made use of the latest zero-points
from STScI, with values of 26.27, 26.26 and 25.96 in F105W, F125W and F160W, respectively. Similarly, {\it Hubble}/ACS imaging data were flat-fielded,
cosmic ray-rejected and charge transfer efficiency (CTE)-corrected with the pipeline {\sc CALACS} (version 2012.2). Exposures were
remapped with {\sc Astrodrizzle} to a pixel scale of $0\farcs03$. The ACS zero points used from an online tool are 25.94 and 25.89 for F814W and F625W, respectively.

The final {\it Hubble} mosaics were astrometrically calibrated to the wider SDSS frame with an overall rms uncertainty, relative to SDSS, of less than 0.05$''$.
This astrometric calibration involved more than 60 galaxies and stars.
The previous Keck/NIRC2 imaging data, due to the limited field of view of 40$''$ in the highest resolution NIRC2 imaging data used for LGS/AO observations,
had large astrometric errors as astrometry was determined based on two bright sources that were also detected in 2MASS. 
Once the HST frames are calibrated, we fixed the astrometry of Keck/NIRC2 image 
with close to 10 fainter sources detected in both WFC3 and NIRC2 images. This astrometric recalibration resulted in a small ($0\farcs1$) shift to the optical
sources relative to the peak PdBI/1.1 mm emission from HFLS3, as can be seen by comparing Figure~1 here with Figure~3 of \citet{Riechers13}.
There is still an overall systematic uncertainty in the relative astrometry between IRAM/PdBI image and {\it Hubble}/Keck images of about $0\farcs1$,
with this value possibly as high as $0\farcs3$ in an extremely unlikely scenario. We account for such a systematic offset in the lens model
by allowing the peak 1.1-mm flux to have an offset from the two lens galaxies with a value as high as $0\farcs3$.

As shown in Figure~1 (left panel), we detect optical emission from more than one galaxy near HFLS3 (galaxies labeled G1 and G2). 
This is similar to what was previously reported with Keck/NIRC2 LGS-AO imaging data, with the southern component (G2) taken to be the rest-frame
optical counterpart to HFLS3 \citep{Riechers13}. If this assumption is correct we expect the southern component to be invisible in the shortest wavelength images, 
as it is a Lyman drop-out   at wavelengths shorter than 8900 \AA. Here, however, we have detected both galaxies in {\it Hubble}/ACS images, 
establishing that G2 is a  galaxy at $z < 5$.
Since these {\it Hubble} observations, we have  reanalyzed the Keck/LRIS spectrum shown in \citet{Riechers13} with $z=2.1$ CIV (1549 \AA) and OIII] (1661, 1666\AA)
emission lines within 1$''$ of HFLS3. We now find some marginal evidence that this
 emission is extended, consistent with the scenario that more than one galaxy may be contributing to the emission lines.
A further confirmation of the redshift of G2 will require additional spectroscopic observations or UV imaging data where $z \sim 2$ galaxies
would be Lyman dropouts. For simplicity, hereafter, we assume that both G1 and G2 are at the same redshift of 2.1. 
The SED modeling we discuss later is consistent with this assumption. 

\section{Rest-frame UV fluxes of HFLS3}

We use the publicly available software {\sc galfit} \citep{Peng02} to model the surface brightness profiles of {\it Hubble}-detected galaxies near HFLS3
and to see if there is any excess emission in WFC3 data relative to the ACS images.
Using {\sc galfit} on the individual {\it Hubble}/ACS and WFC3 frames proved to be difficult because the output models tend to overfit regions of low signal in which HFLS3 
is expected to reside. To remedy this, we stacked the HST/ACS in two bands to increase the signal-to-noise ratio and to model the foreground galaxies in the combined F625W and F814W images.
Under the assumption that the stacked model best represents the two foreground galaxies, we then subtracted the stacked model from individual HST/ACS and WFC3 frames,
with the flux density at each wavelength allowed to vary as an overall normalization in {\sc GALFIT} models.
Any excess in WFC3  relative to ACS would suggest the presence of detectable rest-UV emission from HFLS3. As shown in Figure~1 (right panel) we find
excess emission primarily $0\farcs5$ to the South-West of HFLS3 (labeled R1). We also find some marginal evidence for excess emission near the 1.1-mm peak (labeled R2),
with detection levels between 2.5 to 3.2$\sigma$.
In Table~1 we summarize {\sc galfit} and other intrinsic properties of the two foreground galaxies G1 and G2 as well as the residual
emission R1 and R2, with R2 emission assumed to be from HFLS3.
We also use the latter for a combined UV to far-IR SED modeling with {\sc Magphys} \citep{daCunha08}. We also model the SED of R2 
to determine its photometric redshift and address its association with HFLS3.

\begin{figure}[htb]
\centering
\includegraphics[scale = 0.4]{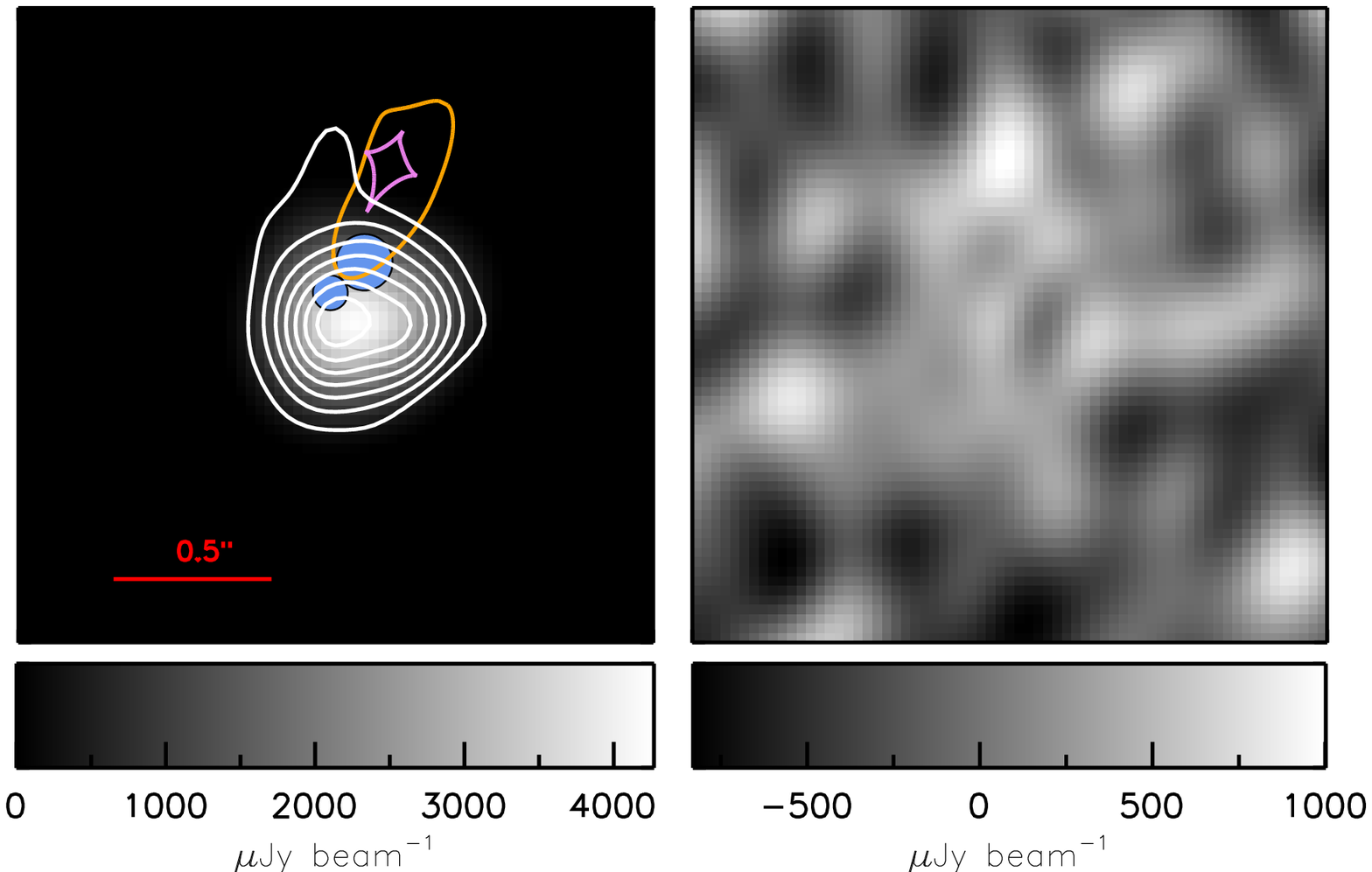}
\includegraphics[scale = 0.4]{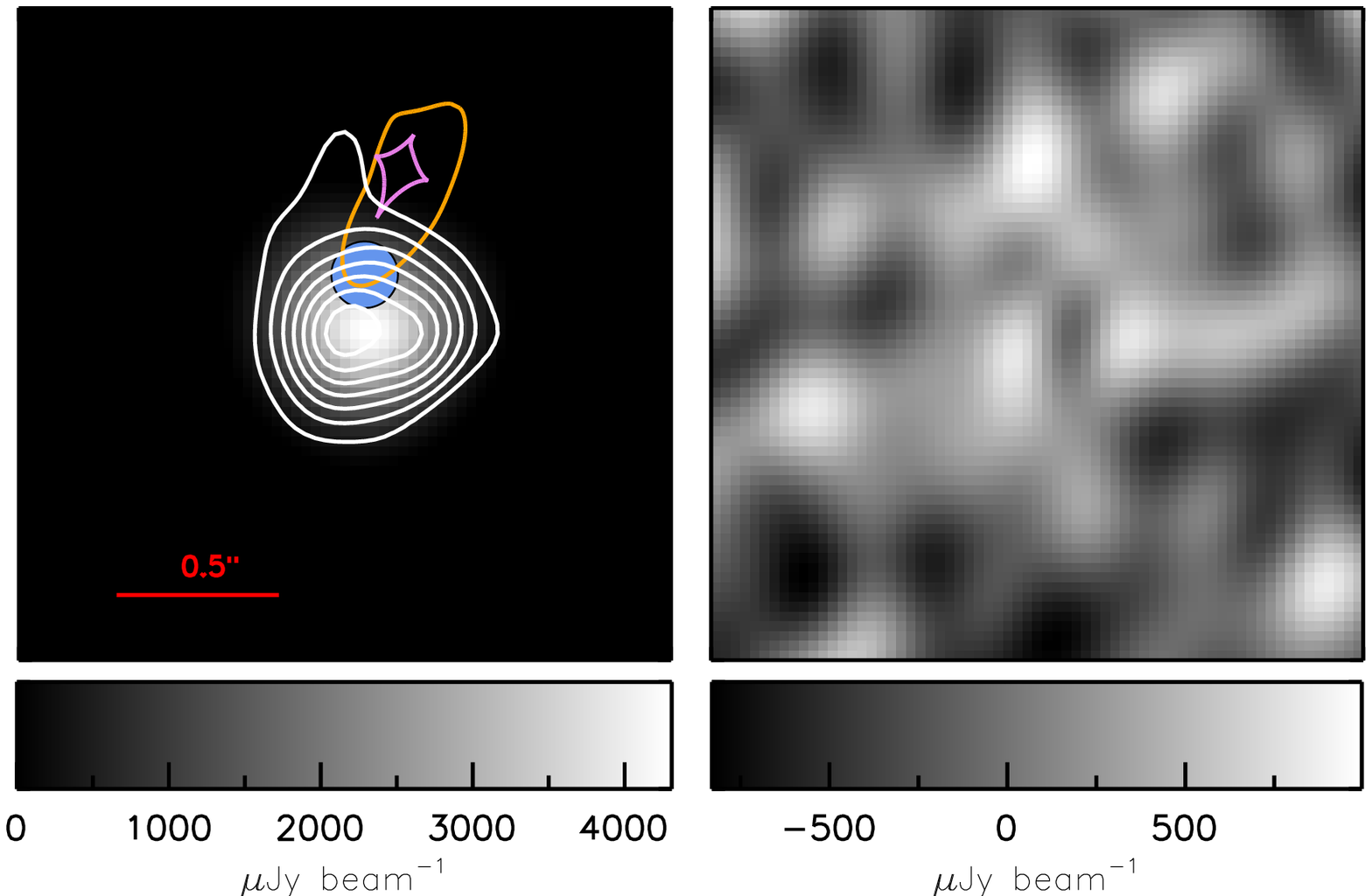}
\caption{\label{fig:lens} 
{\it Top Left:} The source and image plans of HFLS3 under the best-fit lens model with two components for the background source to describe HFLS3. 
In blue we show the two components of HFLS3 in the source plane 
that are gravitationally magnified. The image plane invovles the background grey-scaled
color that  is compared to the measured IRAM/PdBI 1.1-mm continuum emission shown with contours. 
The two lines are the critial line (orange), in the image plane, and the lensing caustic (pink), in the source plane. 
{\it Top Right:} Residual map for the best-fit model showing the difference between observed IRAm/PdBI 1.1-mm continuum emission and the lens model output.
{\it Bottom Left and Right:} The lens model and residual for the case involving HFLS3 described by a single source (shown in blue).
The lines and the residual intensity to the right follow the same as top two panels.
}
\end{figure}

\begin{figure}[htb]
\centerline{
\includegraphics[scale = 0.4]{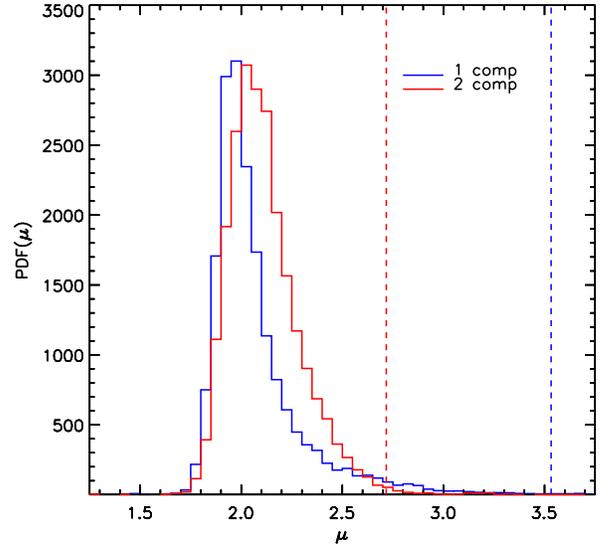}
}
\caption{\label{fig:mag} 
Probability distribution function of the lensing magnification $\mu_{\rm lens}$ at 1.1 mm for HFLS3. We show two scenarios here for the case where
HFLS3 is described by either a single (blue) or double (red) source in the source plane. The vertical lines show the 95\% confidence level upper limit
on the magnification. Note that in both scenarios there is also a strict lower limit for magnification with $\mu_{\rm lens} > 1.6$ at the 95\% confidence level.
The case with $\mu_{\rm lens}=1$ is rejected at $> 6\sigma$ in both cases.
}
\end{figure}

\begin{figure}[htb]
\includegraphics[scale = 0.5]{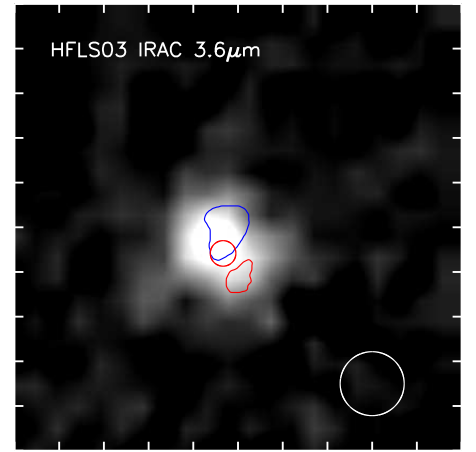}
\caption{\label{fig:galfitirac} 
 10$\times$ 10$''$  {\it Spitzer}/IRAC 3.6 $\mu$m image from Riechers et al. (2013) 
showing a detected source. Each tick mark represents 2$''$. The contours on the intensity scale show the region of G1 and G2 (blue)
and R1 and R2 (red). The IRAC PSF (FWHM = 1.5$''$) is marked with a white circle on the bottom right.
Note that the total flux density measured in IRAC is blended to multiple sources. We are able to measure the total flux density from G1 and G2.
The residual after removing G1+G2 flux density is assumed to be from the new source R1 when attempting to determine
its photometric redshift (Figure~5).
}
\end{figure}

\section{Lens Modeling}

We use the publicly available software {\sc gravlens} \citep{Keeton01} to generate the lens model. 
As the background source is not multiply-imaged, and remains undetected in the rest-frame optical, we measure the goodness of fit
of the model using the highest resolution 
IRAM/PdBI 1.1-mm continuum emission map from \citet{Riechers13}. This map is currently our highest resolution view of HFLS3, and the source is resolved in these data. 
The magnification factor we determine here with lens modeling is the value for the mm-wave continuum emission.
It could be that HFLS3 will be subject to differential magnification, with different emission components within the galaxy
subject to different magnification factors (e.g, \citealt{Serjeant12,Hezaveh12}). This is especially true if the different components
associated with HFLS, such as dust, gas, and stellar mass, have peak intensitities that are offset from each other, as in the case of a complex
merging galaxy system.

To simplify the lens modeling, we use singular isothermal ellipsoidal (SIE) models to fit for the Einstein radius and positions of the two lens galaxies. 
The position angles and ellipticities for G1 and G2 are fixed to the  values derived from profile fitting using {\sc galfit}, but their masses are allowed to vary freely.
The relative positions of G1 and G2 are also kept fixed to {\it Hubble}/ACS-stack measurements, though we do allow the optical galaxy
positions to vary relative to the peak location of the 1.1-mm continuum emission.
For the source plane description of HFLS3, we considered two options: a single source for HFLS3; or a two component model for HFLS3. The latter is motivated by the
fact that the highest resolution [CII] line emission may involve two velocity components separated by about 400~km~s$^{-1}$ \citep{Riechers13}. 
In both these cases, the background source(s) is/are modeled with free parameters for the positions and
effective radii. For simplicity, we assume Gaussian circular profiles with a fixed Sersic index of 0.5. 
The effective radii in the source plane  are allowed to vary in the range of 0.005$''$ to 2.0$''$, with the upper end at a value
higher than the measured size of the 1.1-mm continuum emission in the PdBI image.
In the case of the two component model, the flux ratio between the two background components is also left as a free parameter. 
Hence, the lens model fits for a total of five free parameters for the case with one component for HFLS3 or nine free parameters for the case with two components.
We take this two component model for the background source as a default
model here, though our conclusions do not change if we adopt the single component model. 

In the lens modeling procedure, we output a lensed image as would be observed at 1.1-mm. However, to compare with the data, we convolve
that image with the PdBI beam before calculating the $\chi^{2}$ value. This process is iterated 
over a wide parameter space using the {\sc IDL} routine {\sc amoeba$\_$sa}, which uses a downhill simplex algorithm and simulated annealing to perform multidimensional 
minimization. We  use a circle of a radius $1{\farcs}5$ centered on the peak pixel value of the PdBI 1.1-mm image to measure parameter errors from uncorrelated noise.
For each iteration, the 1.1-mm magnification factor we quote, $\mu_{\rm lens}$, is calculated by simply summing up all the pixel values in 
the image plane and dividing it by the sum of the pixel values in the source plane. 

Figure~2 shows the best-fit model for the two scenarios with one and two components. In the case of two components,
we determine $\mu_{\rm lens}=2.2 \pm 0.3$. The two components have effective radii of $0.5 \pm 0.1$ kpc and $0.3 \pm 0.1$.  
The best-fit model has $\chi^2$ and number of degrees of freedom (Ndof) values of 9929 and  7835, respectively.
For reference, the model with a single component for the background source has $\mu_{\rm lens}=2.0^{+0.9}_{-0.1}$,
an effective radius of  $0.6 \pm 0.1$ kpc, and $\chi^2$/Ndof of 100552/7839.  The lensing masses are 
$M_{\rm lens}= 1.2^{+6.4}_{-0.2} \times 10^9$ M$_{\sun}$ and $1.2^{+0.2}_{-0.1} \times 10^{10}$ M$_{\sun}$ for G1 and G2, respectively
for the two component model. We find masses consistent within these errors for the case when HFLS3 is described by a single component.
The lensing model is mostly sensitive to the mass of G2, while G1, the galaxy fatherst from HFLS3, remains as a minor contribution to the
lens model. Therefore the mass of G1 is less constrained in the lens model.  The best-fit Einstein radius for G1 that we find with
the value of $0{\farcs}05^{+0\farcs06}_{-0\farcs04}$ is barely above the lower value of $0\farcs01$ for the Einstein radius that we placed on the parameter ranges.
We emphasize that the lens model presented here does not require the presence of two lenses in the foreground  or two sources in the background
to fit the data.

Note that we have assumed the redshifts of G1 and G2 are the same in the lens model. The lens magnifications discussed here
are insensitive to the exact assumption related to the redshifts of G1 and G2 as a change in redshift is degenerate with their lensing masses. Our lens model also
assumes a single lensing plane and do not account for multiple-plane lensing if G1 and G2 are at two different redshifts.
If the two galaxies are indeed at the same redshift, they could be part of a galaxy group. The mass we have determined
then could have a contribution from the group potential and will be higher than the value implied by the stellar mass 
of these galaxies. Our lens models do show some evidence for such a possibiity, but due to overall uncertainties in the stellar
mass from SED fits, we cannot reliably confirm this with current data.

In addition to the best-fit lens model, we are also able to place a reliable upper limit on the lensing magnification of HFLS3 at 1.1 mm.
This is simply based on the fact that we have not detected a counter image to HFLS3, while large magnification factors usually result in
image multiplication leading to a detectable counter image. Using the same modeling procedure as described above, 
and allowing for the model to vary over all ranges and including a relative astrometry as high as $0\farcs3$ between 1.1-mm image and lens locations,
we constructed the probability distribution function (PDF)
of magnification for the two cases involving one and two source components to described HFLS3 in the source plane. In Figure~3 we show the histogram where we highlight
the 95\% confidence level upper limit on magnification such that $\mu_{\rm lens} < 3.5$ and 2.7 for  the two cases with one and two components, respectively. The probability distribution functions
also show that there is a strict lower limit to magnification. The case with $\mu_{\rm lens}=1$, where HFLS3 is unlensed, is ruled out 
at more than 6 $\sigma$. This is simply because of the fact that
even a very small mass for G2, the galaxy closest to HFLS3, will result in some lensing magnification of HFLS3.

\begin{figure}[htb]
\centerline{
\includegraphics[scale = 0.45]{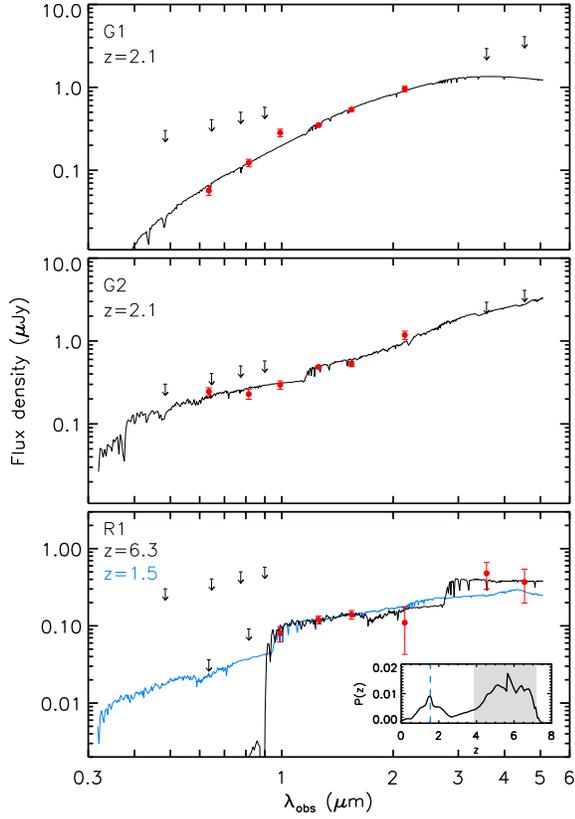}
}
\caption{\label{fig:hyperz} 
SEDs and best-fit {\sc Hyperz} models for optical to IR SEDs of G1  (top), G2 (middle), and R1 (bottom). For G1 and G2 we assume  the optical redshift of 2.1,
though we have yet to establish if the measured optical redshift applies to either G1 or G2, or both. For R1, we allow the redshift to vary as
part of the SED models, and the  probability distribution function for the photometric redshift is shown in the inset to the bottom right of the panel.
We find two solutions with one at high redshift consistent with $z \sim 6$ and a second, involving dusty galaxy templates, at $z \sim 1.3-2.3$. The $\chi^2$ values for the best-fit SEDs are 12.8 and 12.2 for G1 with the number of  degrees of freedom (Ndof) at 11. For R1, the two SEDs show have $\chi^2$ values of 5.4 ($z \sim 6$) and 6.1 ($z\sim 1.5$) with  Ndof of 9.
}
\end{figure}

\begin{figure}[htb]
\centerline{
\includegraphics[scale = 0.5]{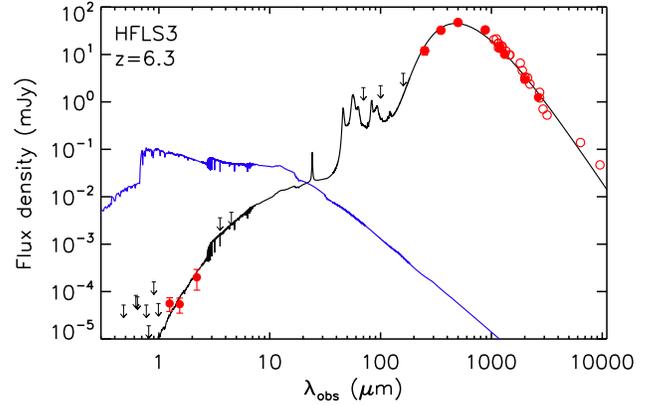}
}
\caption{\label{fig:magphys} 
{\sc Magphys} (da Cunha et al. 2008) 
best-fit SED model of HFLS3 from rest-UV to far-infrared over four decades in wavelength (with a reduced $\chi^2$ value of 1.6). The blue line shows the unobscured template.
For UV to sub-mm SED modeling, we make use of the far-IR/sub-mm data shown by solid symbols. Other measurements shown with open symbols
involve non-standard bands that are not part of the filter and bandpass table of {\sc Magphys}. They were obtained as part of
continuum measurements during atomic and molecular line measurements of HFLS3 with ground-based interferometers.
}
\end{figure}

\section{SED Modeling}

We carry out SED modeling of both foreground and background galaxies
using a combination of {\sc Hyperz}\footnote{v12.2 available from http://userpages.irap.omp.eu/$\sim$rpello/newhyperz/} 
\citep{Bolzonella00} and {\sc Magphys} \citep{daCunha08}. For SED modeling involving G1 and G2, we fix the redshift to the value determined
from optical spectroscopy (Figure~5). We make use of \citet{Bruzual03} SED templates with a combination of single burst, constant, and 
exponentially declining star-fromation history ($\tau$-models), with $\tau$ fixed at 1 Gyr (``E'') and 5 Gyr (``Sb'').
Internal reddening is included using the \citet{Calzetti00} extinction law, and allowing $A_V=0$ to 5 magnitudes in steps of 0.2. We also make use of
the default Lyman-$\alpha$ forest following the prescription from \citep{Madau95}. Given the parameters from the SED modeling (SED type, A$_V$, age etc),
we then make use of the \citet{Bruzual03} models to calculate the H-band mass-to-light ratio assuming a Chabrier IMF \citep{Chabrier03}.
In the case of SED fits, we note that quoted
 error bars are the formal uncertainties and do not include systematic effects. In most of our modeling cases, it could very well be that
uncertain systematics, such as on the choice of SED templates, dominate the error budget.

For G1 and G2, we find stellar masses of $8 \times 10^{8}$ and $1 \times 10^{10}$ M$_{\sun}$, respectively. The ratio of stellar-to-lensing mass for
two lensing galaxies at $z=2.1$ ranges from 0.66 to 0.85.   We find that significant dust attenuation is present in the northern galaxy G1,
 with $A_V \sim 3.4$ mag. We do not find mm-wave emission from that galaxy in our deep 1.1~mm interferometric 
continuum emission data, ruling out a sub-mm bright 
dusty galaxy at this location. It could be that G1 is blended with another galaxy or that our assumption of 2.1 for the redshift is invalid.
As our conclusions related to the physical properties of the two foreground galaxies depends on their redshifts, we caution 
that the properties of these galaxies not be overly-interpreted.
Further deep imaging and spectroscopy should resolve some of the remaining puzzles in the data.

Given the nature and rarity of sources such as HFLS3 at $z > 5$, it is useful to address the extent to which 
the recent {\it Hubble}/ACS and WFC3 imaging changes the
underlying properties of this dusty, starburst galaxy. In order to establish the rest-frame UV fluxes of HFLS3, we made use of the ACS-based models
of the two foreground galaxies to search for excess emission in the longer wavelength data. Note that in ACS, 
$z > 6$ emission should not appear, since
as at those redshifts galaxies will be dropping out of the band due to the Lyman limit.
Using the ACS models on WFC3 data, we found marginal evidence, at around 3.5$\sigma$, 
for  rest-frame UV emission at the PdBI 1.1-mm emission position
in F125W (region marked as R2 in Figure~1 right panel). At the same location we also found 2.8$\sigma$ residual emission in 
both the F160W and Keck/NIRC2 images.
We consider these flux densities to be the rest-UV emission from HFLS3 itself. 

Unfortunately due to blending in the $\sim2''$ PSF, we are not able to deconvolve
the existing {\it Spitzer}/IRAC data to precisely determine the rest-frame optical flux densities of G1, G2 and HFLS3 separately (Figure~4). The IRAC flux densities 
for HFLS3 reported in \citet{Riechers13}  made use of a {\sc Galfit} model for G2 with Keck/NIRC2 image to deblend its contribution from the total.
The residual flux densities are then those corresponding to G1 and HFLS3 in IRAC data and the total residual was assumed to be those
of HFLS3, under the assumption that G1 is the near-IR counterpart of HFLS3.  However, as discussed above, such an assumption no longer applies.
Through the {\sc Galfit} model from {\it Hubble}/ACS data we are able to extract the total G1+G2 IRAC flux densities, but are not able to separate that
total to each of the two galaxies. Thus in Figure~5 we show IRAC flux densities as upper limits for G1 and G2. We detect a residual after removing G1+G2
and in Figure~5 we assume that residual corresponds to R1. A fraction of that residual could also be from HFLS3 at the region marked as R2 in Figure~1 (right panel).
We find that even such an upper limit is subject to assumptions related to {\sc Galfit}  modeling in IRAC images, where multiple  components exist within a single IRAC PSF.
Thus, in Figure~6 we simply use the total flux density measured with IRAC as a conservative upper limit on the flux density of HFLS3 at 3.6 and 4.5 $\mu$m. 

With the rest-UV fluxes for HFLS3 determined with {\it Hubble}/WFC3 and Keck/NIRC2 data fluxes, we cover four orders of magnitude in wavelength from rest-frame UV to far-infrared (Fig.~4).
This SED of HFLS3 is fitted using {\sc Magphys}, where models are calibrated to reproduce
ultraviolet-to-infrared SEDs of local, purely star-forming Ultra Luminous Infrared Galaxies (ULIRGS; 10$^{12} < L_{\rm IR}/{\rm L}_{\odot} < 10^{13}$).
Such models, however, are based on the assumption that dust and stars are in a fully-mixed medium. Massive, dusty starbursts at $z > 2$
may not follow such mixing with differential obscuration causing biases in the combined UV to radio SED. For example,
regions that are bright in the rest-frame optical may only be a small fraction of the regions that are bright in the far-infrared and sub-mm wavelengths.
The use of {\sc Magphys} to model such complex galaxies may result in biased estimates of the physical parameters, but in the absence of
other methods to study the combined SED, we have decided to use {\sc Magphys} here with appropriate cautions.

The SED model assumes a \citet{Chabrier03} initial mass function (IMF) that has a cutoff below 0.1 M$_{\odot}$ and above 100 M$_{\odot}$; 
using a Salpeter IMF instead gives stellar masses that are a factor of $\sim$ 1.7 to 1.8 larger. With {\sc Magphys}-based SED models, we find that 
HFLS3, with rest-UV fluxes in the region marked as R2, shows significant dust attenuation with $A_V \sim 3.6$ mag. Such attenuation is consistent with 
$z\sim 2$ ULIRGs and SMGs (e.g., \citealt{Smail04,Chapman05,Geach07,Swinbank10,Wardlow11,Hainline11,LoFaro13}). 
The best-fit {\sc Magphys} SED model is shown in Figure~6. The fit is dominated by the far-IR/sub-mm data and the overall fit has a reduced 
$\chi^2$ value of 1.6.

\section{Discussion}
\label{sec:discussion}

The {\sc Magphys} SED models of HFLS3 described above lead to SFR, dust mass, stellar mass among other properties. 
As outlined in \citet{Riechers13}, instantaneous SFR, using the FIR luminosity and assuming a \citet{Chabrier03} IMF to scale the \citet{Kennicutt98} relation,
is $\sim$ 2900  M$_{\odot}$ yr$^{-1}$.  Using the {\sc Magphys} SED model, we find that the apparent SFR, averaged over the last 100 Myr, to be 1450 $\pm$ 100 M$_{\odot}$ yr$^{-1}$.
Note that these SFRs must be corrected down by the factor $\mu_{\rm lens}$ to account for lensing magnification. With our preferred best-fit correction factor of $2.2 \pm 0.3$ for the
model involving two components to describe 1.1-mm emission from HFLS3,  the instantaneous and 100-Myr averaged SFRs are
$\sim$ 1300  M$_{\odot}$ yr$^{-1}$ and $\sim 660$ M$_{\odot}$ yr$^{-1}$, respectively. The two are different as
the \citet{Kennicutt98} relation assumes a bolometric luminosity of a constant star-formation lasting over 100 Myr
emitted in the infrared \citep{Kennicutt98,LeithererHeckman95}. For a constant star-formation, bolometric luminosity after the
first 10 Myr evolves relatively slowly as the rate of birth and death of massive stars that dominate the bolometric luminosity reach
a steady state. For starbursting galaxies, however, the SFR is likely changing rapidly over the 100 Myr time interval
and we may be observing the galaxy at the peak of the SFR. Such a possibility then naturally explains why the
instantaneous SFR is a factor of two higher than the SFR averaged over the last 100 Myr. 

We can also place a strict lower limit on the SFRs using
the 95\% confidence level upper limit on lensing magnification. This leads to values of $>$ 780 M$_{\odot}$ yr$^{-1}$ and 390 M$_{\odot}$ yr$^{-1}$ for
instantaneous and 100-Myr averaged SFRs, respectively.  This revision of the SFR to a lower value 
is consistent with a similar revision to the SFR of $z =5.3$ SMG AzTEC-3 \citep{Capak11}. While the total IR luminosity implies a SFR of 
1800 M$_{\odot}$ yr$^{-1}$ \citep{Capak11,Riechers10}, 
SED modeling of the fluxes with population synthesis models have shown the SFR, averged over the last 100 Myr, to be as low as 500 M$_{\odot}$ yr$^{-1}$ \citep{Dwek11}.
Our SED models also show that the age of the oldest stars in HFLS3 is around 200 Myr, suggesting that HFLS3 started assembling its stars at a redshift of
$\sim 8$, during the epoch of reionization.

Using the far-IR/sub-mm SED and the standard assumptions used in {\sc Magphys},  and correcting for magnification,
the dust mass of HFLS3 is $\sim 3 \times 10^8$ M$_{\odot}$, with a lower limit at  $2 \times 10^8$ M$_{\odot}$. The ISM includes two components with dust 
temperatures of 24 $\pm$2 and 50 $\pm$ 2 K.
The best-fit SED model is such that $>$ 90\% of the dust mass is in the warm phase, contrary to low-redshift star-forming galaxies
that have a lower ratio. Such a high ratio for HFLS3 establishes that most of the dust is associated with star-bursting clumps and not the diffuse
cirrus.  The implied dust temperature of the cold phase  component is comparable to the CMB temperature at $z=6.3$, suggesting that the extended cirrus
of this galaxy may be in radiative equillibrium with the CMB. 
Using the \citet{Chabrier03} IMF, with parameters derived again from the SED fits using {\sc Magphys}, and with lensing magnification included, 
we find that HFLS3 has a stellar mass of about $5\times 10^{10}$ M$_{\odot}$. 
This stellar mass, however, is highly uncertain as it is based on just three detections at the rest-frame UV wavelengths. And in all of these cases,
the detections are at the level of 3$\sigma$.  Furthermore, we have assumed that the magnification factor derived with 1.1-mm continuum map also applies for the rest-frame UV emission form which
the stellar mass is derived.  Regardless of these uncertainities, we find that HFLS3 has formed a substantial amount of stellar
mass already. Such a high stellar mass is already at the limits allowed by the dynamical mass of HFLS3 reported in \citet{Riechers13}.

While the SED-based stellar mass is uncertain by an order of magnitude once all modeling errors are accounted for, the 
dust mass of HFLS3 with a value $\sim 3 \times 10^8$ M$_{\odot}$ provides an additional constraint on the  stellar mass of HFLS3.
This comes from models related to the dust formation mechanisms in massive starbursts where
core collapse supernovae (CCSNe) are expected to be the origin of the bulk of the elements that formed the dust.  The
contribution of low mass stars to the refractory elements is negligible in  a young galaxy such as HFLS3.  
Thus, the total number of CCSNe that exploded in the galaxy dictates the maximum dust mass.  Following the
arguments in Watson et~al.  (2014, in preparation), from an observed dust mass, we can infer the minimum number of supernovae that occurred and for a
particular initial mass function, the resulting lower bound on the stellar mass.  The simplest and most robust way to make such an estimate on the stellar mass 
is to work from observations.  SN\,1987A is close to the mass-integrated mean CCSN mass
for most IMFs and is the best-observed CCSN remnant known. Assuming SN\,1987A as a good mass-weighted mean for the dust production, and using the preferred value of a carbonaceous and 
silicate grain mix of $0.6$ to $0.7$\,M$_{\odot}$ \citep{Matsuura11,Indebetouw14}, we can infer that at least $2\times10^{8}$\,M$_{\odot}$ CCSNe exploded in
HFLS3 to account for the dust mass of $3 \times 10^8$ M$_{\odot}$.  

The stellar-to-dust mass ratio should be around 100 for a Chabrier IMF, and a factor of two larger than this for a Salpeter IMF.  The precise value of this ratio
depends on how CCSNe produce refractory metals as a function of mass.  When considering model uncertainties,
the stellar-to-dust mass ratio is within 20\% of the value quoted above, where the dust masses are tied through the observational pivot point provided by the dust mass
observed in SN\,1987A.  Note that this argument is currently based on the dust mass observed
in SN\,1987A as an indication of the refractory element production, rather than claiming that CCSNe necessarily produce all the dust directly. But since
the dust mass observed is believed to be close to the maximal dust production for this SN
\citep{Indebetouw14}, it is therefore a reasonable reflection of the most dust we could ultimately expect to be produced by the elements
synthesised by SN\,1987A.  Thus, for $3\times10^{8}$\,M$_{\odot}$ mass of
dust in the galaxy, we expect a minimum of $\sim 2\times10^{10}$\,M$_{\odot}$
mass of stars for a Chabrier IMF, and twice this for a Salpeter IMF.  This is comparable to the lensing magnification-corrected stellar mass inferred from {\sc Magphys} at $5\times 10^{10}$ M$_{\odot}$, though we note once again
that this value has a large uncertainty due to various assumptions  and low signal-to-noise ratio of the rest-frame UV measurements.
For a SFR averaged over 100 Myr of about 660\,M$_{\odot}$\,yr$^{-1}$, the above arguments imply a characteristic dust
production time of at least 40\,Myr, assuming a negligible dust destruction during the same period.
This is lower than the suggested lifetime for dust mass assembly in AzTEC-3 of about 200 Myr \citep{Dwek11}. 
While our current estimates are uncertain, the above argument, however, can be strengthened in the future
with more precise measurements of dust and stellar masses to constrain dust production mechanisms at $ z\sim 6$.

We also attempted a SED model with far-IR/sub-mm data points combined with rest-UV fluxes from R1, with peak emission
0.5$''$ to the South-West of HFLS3  (Figure~1). 
This emission is detected in all three WFC3 bands at significances greater tha 6 $\sigma$ in each, although the emission
remains undetected in ACS. The emission, however, is blended in IRAC data with the near-IR emission from the two galaxies (Figure~4).
The {\sc Magphys} fit was considerably poor as there was no consistent SED that can
fit the four orders of magnitude in wavelength from UV to sub-mm in that case with the best-fit case having a reduced $\chi^2$ of greater than 5.
This ruled out a scenario in which HFLS3 sub-mm emission is associated with R1. It also rules out an extreme scenario in which our relative astrometry
between IRAM/PdBI and {\it Hubble} images are wrong such that the near-IR counterpart to HFLS3 is R1. We find that R1 must be a separate source.
The {\sc Hyperz} SED model shown in Figure~5 leads to a photometric redshift for this emission is consistent with a source at $z \sim 6.3$ (Figure~2 bottom panel), though a dusty galaxy SED at $z \sim 2$ is also consistent with this emission. The {\sc Hyperz} fit to the data leads to a stellar mass of  $\sim 1.2 \times 10^{10}$ M$_{\odot}$ for R1 if we assume
the redshift is $z=6.3$, following the $z \sim 6$ photo-z solution. 

We have two possibilities for this new source. It could be part of the  emission associated with a complex galaxy merger system involving HFLS3, especially if HFLS3
starburst is triggered by a merger as is the case for most $z \sim 2$ bright SMGs.
Alternatively, it could be part of the $z \sim 2.1$ foreground structure that is responsible for lensing of HFLS3. If the latter is indeed the case, the region in the foreground
of HFLS3 involves a massive galaxy group, but the magnification upper limit of 3.7 we have derived here is unlikely to be revised higher
as it accounts for a wide variation of model parameters, including to the total lens mass in the foreground. It is far more likely that
R1 is part of the complex merger system associated with HFLS3.

\section{Summary}

Here, we have discussed the rest-frame ultraviolet emission from the starbursting galaxy HFLS3 at a redshift of 6.34.
The recently acquired {\it Hubble}/WFC3 and ACS imaging data show conclusively that the previously identified rest-frame optical counterpart of HFLS3 is
at $z < 6$. We find two galaxies in the foreground leading, to a clear possibility for lensing magnification, though at a level below that needed to form
multiple images.  A lensing model based on the {\it Hubble} imaging data then leads to
a magnification factor for the mm-wave continuum emission of   $2.2 \pm 0.3$, with a strict upper limit of 3.7 at the 95\% confidence level. 
The scenario involving no lensing is ruled out at more than 6 $\sigma$ confidence level.  
Using models for the rest-frame UV to far-IR spectral energy distribution
we determine the instantaneous SFR, 100 Myr-averaged SFR, dust, and stellar masses of HFLS3
to be 1320 M$_{\sun}$ yr$^{-1}$, 660 M$_{\odot}$ yr$^{-1}$,  $3 \times 10^8$ M$_{\odot}$, and $5 \times 10^{10}$ M$_{\odot}$, respectively,
with large uncertainties especially on the stellar mass of HFLS3. The properties of HFLS3 suggest a galaxy that has intrinsic properties that are roughly consistent
with  $z =5.3$ SMG AzTEC-3, but there are also differences resulting from the higher dust and stellar mass of HFLS3.

Galaxies with sub-mm colors similar to HFLS3 have been now identified in SPIRE data, leading
to the possibility that detailed statistical studies on massive, dusty, star-bursts during reionization will become feasible with future facilities \citep{Dowell14}.
While statistical studies will be necessary to address fundamental questions regarding how such massive, metal-rich, starbursting  galaxies could form 800 Myr after
the Big Bang, detailed studies of individual galaxies are also useful to address whether the astrophysics that govern massive starbursts during reionization are
similar to those in $z \sim 2$ sub-millimeter galaxies.

\acknowledgments
Financial support for this work was provided by NASA
through grant HST-GO-13045 from the Space Telescope Science Institute, which is operated
by Associated Universities for Research in Astronomy,
Inc., under NASA contract NAS 5-26555.
Additional support for AC, WO, JC, JLW, and CMC was from NSF with AST-1313319.
We thank E. da Cunha for help with {\sc Magphys}. 
Dark Cosmology Centre is funded by the Danish National Research Foundation (JLW and DW).
SO acknowledges support from the Science and Technology Facilities Council [grant number ST/I000976/1].
SPIRE has been developed by a consortium of institutes led
by Cardiff Univ. (UK) and including: Univ. Lethbridge (Canada);
NAOC (China); CEA, LAM (France); IFSI, Univ. Padua (Italy);
IAC (Spain); Stockholm Observatory (Sweden); Imperial College
London, RAL, UCL-MSSL, UKATC, Univ. Sussex (UK); and Caltech,
JPL, NHSC, Univ. Colorado (USA). This development has been
supported by national funding agencies: CSA (Canada); NAOC
(China); CEA, CNES, CNRS (France); ASI (Italy); MCINN (Spain);
SNSB (Sweden); STFC, UKSA (UK); and NASA (USA).
The data presented in this paper will be released through the {\em Herschel} Database in Marseille HeDaM ({hedam.oamp.fr/HerMES})

\clearpage

\begin{deluxetable}{lcl}
\vspace{-10mm}
\tablewidth{0pt}
\tablecaption{IR Properties of HFLS3 and Near-by Galaxies
 \label{tab:fls}}
\tablehead{
\colhead{Quantity} & \colhead{Value} & \colhead{Ref}}
\startdata
\multicolumn{3}{c}{G1}\\
\hline
RA  &  $17:06:47.80$ & ACS $I_{814}$-band\\
Dec &  $+58:46:24.33$ & ACS $I_{814}$-band\\
Redshift & 2.019 & Riechers et al. 2013 \\
ACS/F625W   & $27.01\pm0.14$ (AB mag) &   Photometry \\
ACS/F814W   & $26.17\pm0.12$ (AB mag) &   Photometry\\
WFC3/F105W   & $25.27\pm0.12$  (AB mag) &   Photometry \\
WFC3/F125W   & $25.27\pm0.04$ (AB mag) &   Photometry \\
WFC3/F160W   & $24.57\pm0.09$ (AB mag) &   Photometry \\
NIRC2/K$_{\rm s}$-band & $23.94\pm0.04$ (AB mag) & Photometry \\
$R_{\rm E}$ &  $ 0\farcs05_{-0\farcs01}^{+0\farcs06}$  & lens model\\
$M_{\rm E}$ &   $(1.2_{-0.2}^{+6.4})\times10^{9}$ M$_{\odot}$ & lens model\\
$R_{\rm e}$ & $0\farcs9\pm0\farcs3$& {\tt GALFIT} \\
      &   $7.1\pm2.3 $ kpc &    \\
$\epsilon$ & $0.48\pm 0.02$ & {\tt GALFIT} \\
PA$_{\rm d}$ & $(88\pm2)^{\circ}$ & {\tt GALFIT}\\
$n$ (S\'ersic) &  $4.3\pm0.8$ &  {\tt GALFIT} \\
$M_{\star}$ & $\sim 8  \times10^{8}$ M$_{\odot}$ & SED ({\sc Hyperz})\\
$A_V$ & $\sim 3.4$ mag & SED ({\sc Hyperz})\\
$L_V$ (extinction corrected) & $\sim 3\times10^{11}$ L$_{\odot}$ & SED ({\sc Hyperz})\\
\hline
\multicolumn{3}{c}{G2}\\
\hline
RA  &  $17:06:47.77$ & ACS $I_{814}$-band\\
Dec &  $+58:46:23.95$ & ACS $I_{814}$-band\\
ACS/F625W   &  $25.42\pm0.13$ (AB mag) &   Photometry \\
ACS/F814W   &  $25.50\pm0.16$ (AB mag) &   Photometry\\
WFC3/F105W   &  $25.22\pm0.13$ (AB mag) &   Photometry \\
WFC3/F125W   &  $24.68\pm0.05$ (AB mag) &   Photometry \\
WFC3/F160W   &  $24.59\pm  0.13$ (AB mag) &   Photometry \\
NIRC2/K$_{\rm s}$-band &  $ 23.72\pm0.09$ (AB mag) & Photometry \\
$R_{\rm E}$ & $ 0\farcs15_{-0\farcs01}^{+0\farcs02}$ & lens model\\
$M_{\rm E}$ &   $(1.2_{-0.1}^{+0.2})\times10^{10}$ M$_{\odot}$ & lens model\\
$R_{\rm e}$ & $0.34\pm0.01''$ & {\tt GALFIT} \\
      &  $2.8\pm0.1$ kpc &    \\
$\epsilon$ & $0.63 \pm 0.01$ & {\tt GALFIT} \\
PA$_{\rm d}$ & $(-30\pm1)^{\circ}$ & {\tt GALFIT}\\
$n$ (S\'ersic) &  $0.98\pm0.03$ &  {\tt GALFIT} \\
$M_{\star}$ &  $1\times10^{10}$ M$_{\odot}$ & SED ({\sc Hyperz})\\
$A_V$ & 1.20 mag & SED ({\sc Hyperz})\\
$L_V$ (extinction corrected) & $4\times10^{10}$ L$_{\odot}$ & SED ({\sc Hyperz})\\
\hline
\multicolumn{3}{c}{HFLS3}\\
\hline
RA  &  $17:06:47.80$ & Riechers et al. 2013 \\
Dec & $+58:46:23.51$& Riechers et al. 2013 \\
Redshift & $6.3369 \pm 0.0009$ & Riechers et al. 2013 \\
ACS/F625W   & $>27.01$ (AB mag) &   Photometry \\
ACS/F814W   & $>28.20$ (AB mag) &   Photometry\\
WFC3/F105W   & $>27.58$ (AB mag) &   Photometry \\
WFC3/F125W   & $27.02\pm0.35$ (AB mag) &   Photometry \\
WFC3/F160W   & $27.06\pm0.38$  &   Photometry \\
NIRC2/K$_{\rm s}$-band & $25.64\pm0.50$& Photometry \\
$\mu_{\rm lens}$ & $2.2 \pm 0.3$ & two-component model\\
$\Theta_{s1}$ &  $0\farcs5 \pm 0\farcs1$  &  component 1\\
              &  $2.6 \pm 0.7$ kpc & \\
$\Theta_{s2}$ &  $0\farcs3^{+0\farcs2}_{-0\farcs1}$  &  component 2\\
              &  $1.6^{+1.2}_{-0.6} $  kpc & \\
$\frac{F_{2}}{F_{1}}$ & $0.3^{+0.4}_{-0.2}$  & Flux ratio \\
$M_{\rm dust}$ & $3 \times 10^8$ M$_{\odot}$ & SED ({\sc magphys})\\
SFR$_{\rm int}$ & 1320 M$_{\sun}$ yr$^{-1}$ & Kenicutt (1998)\\
$\langle {\rm SFR} \rangle_{100 Myr}$ & $654^{+104}_{-90}$ M$_{\odot}$/yr & SED ({\sc Magphys})\\
$M_{\star}$ &   $\sim 5 \times10^{10}$M$_{\odot}$ & SED ({\sc Magphys})\\
$A_V$ & $3.6$ mag& SED ({\sc magphys})\\
$L_V$ (extinction corrected) & $\sim 4\times10^{12}$L$_{\odot}$ & SED ({\sc Magphys}) \\
\hline
\multicolumn{3}{c}{R1}\\
\hline
RA  &  $17:06:47.76$ & WFC3 $H_{160}$-band\\
Dec &  $+58:46:22.87$ & WFC3 $H_{160}$-band\\
ACS/F625W   & $>27.01$(AB mag) &   Photometry \\
ACS/F814W   &  $>26.85$ (AB mag) &   Photometry\\
WFC3/F105W   & $26.68\pm0.28$ (AB mag) &   Photometry \\
WFC3/F125W   & $26.20\pm0.14$ (AB mag) &   Photometry \\
WFC3/F160W   & $ 26.03\pm0.15$ (AB mag) &   Photometry \\
NIRC2/K$_{\rm s}$-band & $ 26.30\pm 0.92$ (AB mag) & Photometry \\
\enddata
\tablecomments{
We assume $z_{\rm G1}=z_{\rm G2}$. For non-detections, flux density upper limits are given at 3$\sigma$.
For HFLS3 flux densities are not corrected for lensing magnification, but physical properties are assuming
$\mu=2.2$. We do not list IRAC flux densities due to large uncertainities in deblending and the separation of total flux density to the four source components.
}
\end{deluxetable}





\end{document}